\documentclass[aps,prl,twocolumn,showpacs,groupedaddress]{revtex4}

\usepackage{graphics}

\begin{document}


\title{Dimensional trend in CePt$_2$In$_7$, Ce-115 compounds, and CeIn$_3$}


\author{Munehisa Matsumoto$^{1}$, Myung Joon Han$^{2}$, Junya Otsuki$^{3}$, Sergey Yu. Savrasov$^{1}$}
\affiliation{$^1$Department of Physics, University of California, Davis, California 95616, USA\\
$^2$Department of Physics, Columbia University, 538 West 120th Street, New York, New York 10027, USA\\
$^3$Department of Physics, Tohoku University, Sendai 980-8578, Japan}


\date{\today}

\begin{abstract}
We present realistic Kondo-lattice simulation results for
the recently-discovered heavy-fermion antiferromagnet CePt$_2$In$_7$
comparing with its three-dimensional counterpart CeIn$_3$ and the
less two-dimensional ones, Ce-115's.
We find
that the distance to the magnetic quantum critical point
is the largest for CeIn$_3$ and the smallest for
Ce-115's, and CePt$_2$In$_7$ falls in between. We argue that the trend
in quasi-two-dimensional materials stems from the frequency dependence
of the hybridization between Cerium $4f$-electrons and the conduction bands.
\end{abstract}

\pacs{71.27.+a, 74.10.+v, 75.40.Mg} 

\maketitle


Recently-discovered CePt$_2$In$_7$~\cite{kurenbaeva_2008,bauer_2010}
has provided a new approach toward the two-dimensional (2D)
limit in CeIn$_3$-derived heavy-fermion material family, among which Ce-115 materials
have been discussed intensively in the past decade. In the hot debates on 115's,
the possible scaling of the superconducting transition temperature $T_{\rm c}$
to $c/a$, where $a$ and $c$ are the lattice constants of the tetragonal
crystal structure, has been discussed~\cite{bauer_2004,dix_2009} and there has been
a hope to raise $T_{\rm c}$ by making a more 2D-like material. In this context $c/a=4.694$
(if $c$ should be taken to be the interlayer distance between Cerium planes,
the scaling parameter should be taken to be $2.347$) in CePt$_2$In$_7$~\cite{kurenbaeva_2008}
looks promising as compared to the typical values of
$c/a\sim 1.6$ in Ce-115's where the highest $T_{\rm c}$'s
have been found among Cerium heavy-fermion
compounds~\cite{hegger_2000,petrovic_2001,petrovic_2001_EPL}.

Theoretically, the superconductivity (SC) with high $T_{\rm c}$'s in strongly-correlated materials
has been discussed to be mediated by magnetic fluctuations~\cite{monthoux_2007}
and the possible mechanism for SC seems to be intimately related to the nearby
magnetic quantum critical point (QCP)~\cite{sachdev_1999}. An empirical
material-designing principle to get high $T_{\rm c}$
would be to make it as close as to QCP and it is desirable for a given
possibly high-$T_{\rm c}$ material to know how close it is located to QCP.
The robustness of the magnetic pairing has been discussed
considering the spatial dimensional effects~\cite{monthoux_2001},
stressing the stronger robustness in the more 2D-like systems even though
care must be taken regarding the details of the electronic structure.
Thus we are motivated to address QCP in CePt$_2$In$_7$ fully
taking into account its electronic structure and
see the trend among the related materials, namely, CeIn$_3$ with $c/a=1$ and Ce-115's with slightly
larger $c/a\simeq 1.6$, to see if any microscopic origin
of $c/a$-scaling of $T_{\rm c}$ could be traced quantitatively to QCP.
In the present work we predict the QCP's for all of these materials to elucidate where exactly
CePt$_2$In$_7$ is located in the neighborhood of QCP.

Recent experiments on CePt$_2$In$_7$~\cite{nick_2010} have shown
that this material is an ideal Kondo lattice with commensurate antiferromagnetism.
It can be a good target for realistic Kondo lattice simulations~\cite{mm_2009} for which
it has been shown that the description works fine in the Kondo limit,
meaning that local $f$-level position, $\epsilon_{f}$, measured
from the Fermi level should be negative and large,
$U+\epsilon_{f}$ should be positive and large
with $U$ being the on-site Coulomb repulsion energy,
and at the same time hybridization $V^2$ should not be too big.
When the valence fluctuations start to dominate
the Kondo lattice simulations does not work and we have to go back to the lattice
simulations based on the Anderson impurity problem~\cite{anderson_1961},
which would increase the computational cost. It has been discussed that Ce-115's,
especially CeRhIn$_5$ and CeIrIn$_5$, have well localized electrons~\cite{fujimori} and
we expect that the comparison between CePt$_2$In$_7$,
CeRhIn$_5$/CeIrIn$_5$, and CeIn$_3$ in our simulations would make sense,
while CeCoIn$_5$ might have to be looked at with some extra care.

\begin{figure}
\scalebox{0.8}{\includegraphics{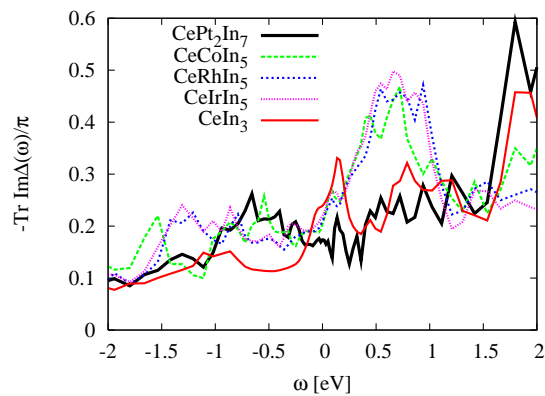}}
\caption{\label{ImDelta} (Color online) Hybridization function calculated
by local density approximation (LDA) with Hubbard-I approximation
for CePt$_2$In$_7$, Ce-115's and CeIn$_3$ using the code described in Ref.~\cite{sergey}.
The trace in the vertical
axis quantity is taken over 14 orbitals of the $4f$-orbital of Cerium.}
\end{figure}
We define our realistic Kondo lattice Hamiltonian starting
with the first-principles electronic-structure
calculation for the Cerium heavy fermion materials.
The local density approximation (LDA) for delocalized $s$-, $p$-, and $d$-conduction
electrons combined with the Hubbard-I approximation~\cite{hubbard_1963}
for localized $f$-electrons
gives the frequency-dependent
hybridization function, $-\Im\Delta(\omega)/\pi$,
between the conduction electrons and the localized $4f$-electrons
as shown in Fig.~\ref{ImDelta}. A part of the trend is already seen at this stage:
we find that the hybridization around the Fermi level, which is most relevant
for Kondo screening~\cite{burdin_2000}, is the strongest for CeIn$_3$ and
the weakest for CePt$_2$In$_7$, and Ce-115's comes in between. This
is in line with the relevance of the off-plane hybridization that has been
discussed for CeIrIn$_5$~\cite{shim_2007}.
Thus obtained data reflects the high-energy physics of the given material and
then we obtain our low-energy Hamiltonian via
the Schrieffer-Wolff transformation~\cite{schrieffer_1966} implemented
in a realistic way~\cite{mm_2009}.
The Kondo coupling
$J_{\rm K}=V^2[1/|\epsilon_{f}|+1/(U_{\rm eff}+\epsilon_{f})]$
incorporates the virtual hopping process $f^1 \rightarrow f^0$
in the first term and in the second term another virtual process $f^1 \rightarrow f^2$
with $U_{\rm eff}=U-J_{\rm Hund}\sim 4$~[eV]
being the effective on-site Coulomb energy
in the $f^2$ multiplet in the Ce atom including the effective Hund coupling
$J_{\rm Hund}\sim 1$~[eV].
The hybridization $V^2$ is taken to be
$\int_{-\infty}^{D}d\omega\,[-\Im\Delta(\omega)/\pi]$,
to take care of the normalization of the density of states of our conduction electrons
with keeping the value of a relevant quantity in Kondo physics,
$J_{\rm K}\rho(\omega)$, where $\rho(\omega)$ is the conduction electron density of states,
to the given realistic value. Note that the conduction band cutoff $D$
matters in the definition of $J_{\rm K}$. Although $-\Im\Delta(\omega)/\pi$ is a decaying
function of $\omega\gg 0$, it does not fall off sufficiently fast due to numerically
imperfect projection onto $f$-states. Thus the cutoff $D$ is implemented to be equal to
the on-site Coulomb energy $D=U=5$ [eV].
The relevant parameters and the outputs of LDA for the Cerium compounds of our
present interest is summarized in Table~\ref{numbers}.
We note that our calculated values of $\epsilon_{f}$ for Ce-115's
are not exactly consistent with the photoemission data
in the literature~\cite{fujimori,koitzsch_2009}.
However both of the calculated values and experimental ones
indicate that all of the materials are deep in the Kondo regime.
The spin-orbit splittings are set to be $0.3$~[eV]
for all of the materials.
The structural parameters to be plugged into LDA are taken from experiments
referring to Ref.~\cite{rmp_2009}.
\begin{table}
\caption{\label{numbers}
The parameters for the investigated materials.
The first two columns are LDA + Hubbard-I results
for the local $f$-level position $\epsilon_{f}$
and the value of the hybridization function on the Fermi level.
The final column
shows the crystal-field splittings in the tetragonal structure.}
\begin{tabular}{ccccc}\hline
 & & $\epsilon_{\rm f}$ [eV] & $-{\rm Tr}\Im\Delta(0)/\pi$ [eV] & $\Delta_{1}$ and $\Delta_{2}$ [meV]\\ \hline
2D & CePt$_2$In$_7$& -1.81  & 0.174 & 8.6, 12.9$^a$ \\ \hline\hline
& CeCoIn$_5$ & -1.97 & 0.205 & 6.8, 25$^b$ \\ \hline
& CeRhIn$_5$ & -1.90 & 0.209 & 5.9, 28.5$^c$ \\ \hline
& CeIrIn$_5$ & -1.95 & 0.220 & 5.26, 25.875$^c$\\ \hline\hline
3D & CeIn$_3$ & -1.72 & 0.239 & 12$^d$ \\ \hline
\end{tabular}
\begin{flushleft}
$^a$ = 100 [K] and 150[K], a rough estimate~\cite{nick_2010_2}.\\
$^b$ Ref.~\cite{latest,christianson_2004}. Another crystal-field scheme in Ref.~\cite{nakatsuji_2002}
gives essentially the same results.\\
$^c$ Ref.~\cite{takeuchi_2001}. The latest crystal-field schemes in
Ref.~\cite{latest} give the values close to these.\\
$^d$ Ref.~\cite{rmp_2009}. This materials has the cubic structure which brings
$\Delta_{1}=\Delta_{2}$.\\
\end{flushleft}
\end{table}

Now we describe how
we solve the realistic Kondo lattice model (KLM) that we have defined
from $-\Im\Delta(\omega)/\pi$
in Fig.~\ref{ImDelta} and Table~\ref{numbers}.
We use dynamical mean-field theory (DMFT)~\cite{georges_1996,gabi_2006} that is formulated
on a local $f$-electron basis~\cite{otsuki_2009_formalism} which
enables us to reach low temperature region
with a modest computational cost and address the QCP in a semi-quantitative way
utilizing state-of-the-art continuous-time quantum Monte Carlo (CT-QMC)
impurity solver~\cite{rubtsov_2005,werner_2006,haule_2007,otsuki_2007}.
We plug-in the realistic crystal-field and spin-orbit splittings as given in Table~\ref{numbers}
in the local $4f$-level in the impurity problem.
Thus our solutions are numerically exact up to the approximation of DMFT.
We look at the magnetic phase transition as a function of temperature and
restore the Doniach phase diagram~\cite{doniach_1977} by varying
the Kondo coupling $J_{\rm K}$  for a given material.
Thus
a magnetic QCP is found on a realistic Doniach phase diagram spanned by the Kondo coupling
and temperature, and the realistic data point
for the given material is picked up for the realistic value of $U=5$~[eV]
and $J_{\rm Hund}=1$~[eV] to estimate its distance to QCP.

\begin{figure}
\scalebox{0.8}{\includegraphics{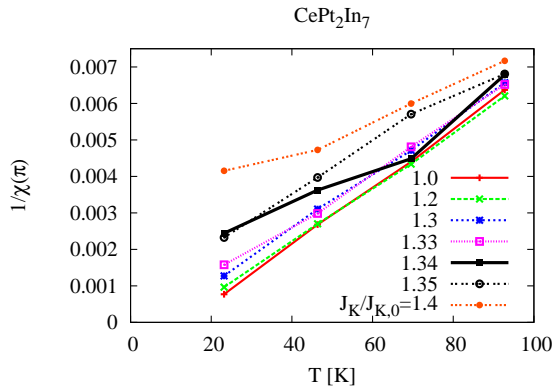}}
\caption{\label{ssus} (Color online) Temperature dependence of the inverse
of the staggered susceptibility in the realistic Kondo lattice models for
CePt$_2$In$_7$ by which we identify that the phase
boundary is located in
$1.33J_{{\rm K},0}< J_{\rm K}<1.34J_{{\rm K},0}$, where $J_{\rm K,0}$ is the
Kondo coupling which corresponds to $U=5$~[eV] and $J_{\rm Hund}=0$.}
\end{figure}

\begin{figure}
\scalebox{0.8}{\includegraphics{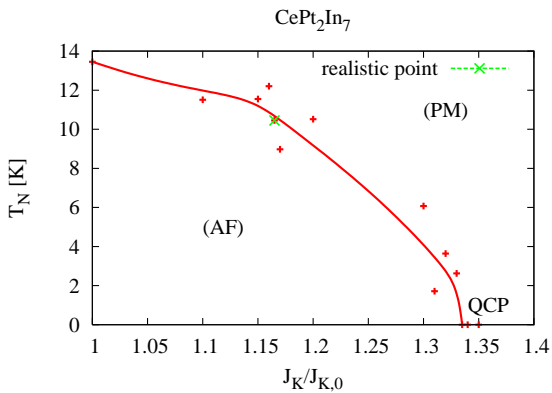}}
\caption{\label{doniach_raw_CePt2In7} (Color online) Restored realistic Doniach phase diagram for CePt$_2$In$_7$
from the data in Fig.~\ref{ssus}. AF is the antiferromagnetic phase and PM is the paramagnetic phase.
The line is a guide to the eye.}
\end{figure}

Now we show how we determine the QCP of CePt$_2$In$_7$.
The calculated temperature dependence of the inverse of
the staggered magnetic susceptibility, $1/\chi(\pi)$, for CePt$_2$In$_7$ is shown in Fig.~\ref{ssus}.
Here we have employed random dispersion approximation~\cite{gebhard_1997}
to estimate the two-particle Green's function by decoupling,
which would enhance our transition temperatures on top of DMFT as will be seen below.
Being consistent with Doniach's picture~\cite{doniach_1977} where the winner of the
competition between the magnetic-ordering energy scale $\propto J_{\rm K}^{2}\rho$ and the
Kondo-screening energy scale $\propto \exp[-1/(J_{\rm K}\rho)]$ interchanges at some finite $J_{\rm K}$,
it is seen that small $J_{\rm K}$'s give a diverging $\chi(\pi)$ at a finite N\'{e}el temperature $T_{\rm N}$
while large $J_{\rm K}$'s give a saturating $\chi(\pi)$ at low temperatures, and some value
of $J_{\rm K}$ in between gives the quantum critical point where $T_{\rm N}$ vanishes.
The N\'{e}el temperatures for smaller $J_{\rm K}$'s are determined by linear extrapolation
of $1/\chi(\pi)$ to the lowest temperature region and thus obtained $T_{\rm N}$ is plotted
against the Kondo coupling in Fig.~\ref{doniach_raw_CePt2In7} in a format of
restored Doniach phase diagram. Here we note that
we just identified what we call QCP by a parameter segment where the
finite N\'{e}el temperature seems to have vanished,
and there is always a possibility that in some smaller parameter segments there is actually
a coexistence region. There is also
a possibility numerically for a first-order phase transition where the N\'{e}el temperature actually jumps
from a finite value to zero at a certain point of $J_{\rm K}$.
We leave the exact characterization of what we also call QCP here for future investigations
and for the moment we would be satisfied with that it looks like QCP practically
in most cases with a very small jump numerically, if any.

\begin{figure}
\scalebox{0.8}{\includegraphics{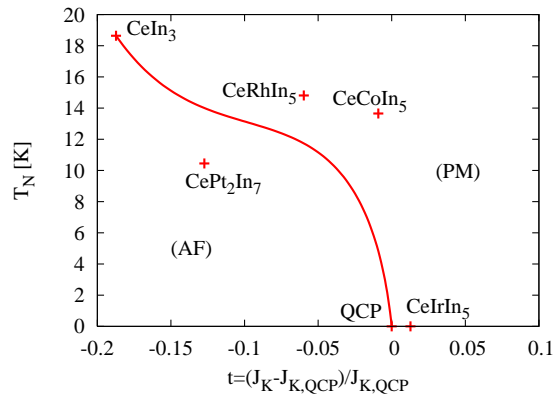}}
\caption{\label{doniach} (Color online) Realistic Doniach phase diagram for CePt$_2$In$_7$,
and Ce-115's together with their parent compound, CeIn$_3$ with the horizontal axis
rescaled to measure the distance to the magnetic QCP. The line is a guide to the eye.
}
\end{figure}

For CePt$_2$In$_7$, the realistic data point is at
$J_{\rm K}=1.165 J_{\rm K,0}$, which corresponds to $U=5$~[eV]
and $J_{\rm Hund}=1$~[eV],
gives the N\'{e}el temperature $T_{\rm N}\simeq 10$ [K] which is larger than the experimental
result $T_{\rm N}=4.5$ [K]~\cite{nick_2010} due to our mean-field argument with DMFT and
the random dispersion approximation~\cite{gebhard_1997} which becomes exact on a lattice with perfect nesting.
The same analyses are applied to the other materials to give the N\'{e}el temperature
$T_{\rm N}=19$~[K] for CeIn$_3$, $15$~[K] for CeRhIn$_5$, and $14$~[K] for CeCoIn$_5$. Again the experimental values
($T_{\rm N}=10.2$~[K] for CeIn$_3$, $3.8$~[K] for CeRhIn$_5$~\cite{hegger_2000,rmp_2009}) comes below these results.
The data for CeCoIn$_5$ is not consistent with  the experimental fact that CeCoIn$_5$
is a non-magnetic heavy fermion material~\cite{petrovic_2001}. However the arrangement
of the materials around QCP reveals that the apparent finite $T_{\rm N}$ actually comes from
being in an immediate proximity to QCP as shown in Fig.~\ref{doniach}.
We use a rescaled Kondo coupling on the horizontal axis in terms of its value right on the QCP,
$t\equiv (J_{\rm K}-J_{\rm K}|_{\rm QCP})/J_{\rm K}|_{\rm QCP}$, to remove a problem
with the Doniach phase diagram that each material has its own energy scales.

Now we inspect the distance to the QCP of CePt$_2$In$_7$ referring
to those of Ce-115's and CeIn$_3$. The cubic parent material CeIn$_3$ is seen to
be most separated from QCP and Ce-115's are found to be concentrated in the neighborhood to QCP,
with CeRhIn$_5$ on the magnetic side and CeIrIn$_5$ on the non-magnetic side. Our numerical
resolution is not sufficient to locate CeCoIn$_5$ in its correct non-magnetic side, but
it is clear that it sufficiently works to estimate the extreme closeness to QCP of Ce-115's.
We note that our calculation scheme might not be as good for CeCoIn$_5$ as for the others
due to the possibly stronger effects of valence fluctuations in this material.
It is seen that when we try to reach the quantum critical point (QCP)
from CeIn$_3$ in the three-dimensional limit, CePt$_2$In$_7$ is located in the midway
toward Ce-115's which are located closest to the QCP.
Seen from QCP, CeIn$_3$ is already close enough
to enable the pressure-driven superconductivity~\cite{mathur_1998}
and CePt$_2$In$_7$ would also have one.
Making a material more 2D indeed helps it to come closer to QCP,
which is reasonable in the general context that the lower
spatial dimensionality would suppress the magnetic long-range order.
However within the 2D-side, the trend among CePt$_2$In$_7$ and Ce-115's is
somewhat non-monotonic. The possible reason is seen in Fig.~\ref{ImDelta},
where a dip around the Fermi level is seen for CePt$_2$In$_7$
which would help to drive it to the magnetic side with the reduced Kondo screening.
On top of the spatial dimensionality, the frequency dependence of the hybridization
seems to introduce the nontrivial trend in this way.

Here we note that precisely speaking we are discussing in the infinite-dimensional
limit with the DMFT, but at least a semi-quantitative trend of $T_{\rm N}$
would be satisfactorily addressed.
We also note that
in the literature
the importance of momentum dependence of the hybridization
has been stressed~\cite{shim_2007,burch_2007,haule_2009} which we have neglected.
What we have seen is that the energy dependence seems to be sufficient at least
to capture the trend in the distance to QCP, thus for the prediction of
magnetically-mediated superconductivity with a possible high $T_{\rm c}$.

In order to reach QCP, it would be interesting to have a material analogous to CePt$_2$In$_7$
but without a big dip in the hybridization around the Fermi level. For that,
the following possible ways for the material designing could help:
1) electronic carrier doping, 2) ascending T in the periodic table
for CeT$_2$In$_7$ to enhance the hybridization,
and 3) shifting T to the left- or the right-hand side on the periodic table
to lift the dip off the Fermi level
to enable the stronger Kondo coupling to drive the material toward the QCP side.

\begin{figure}
\scalebox{0.8}{\includegraphics{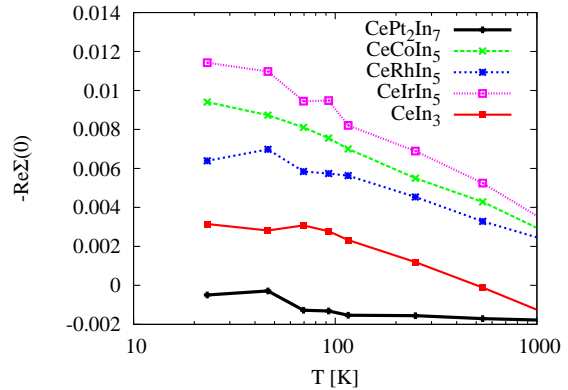}}
\caption{\label{plot_minus_ReSigma} (Color online) Temperature dependence of the real part
of the conduction-electron self-energy for CePt$_2$In$_7$, Cerium-115's and CeIn$_3$.}
\end{figure}

Finally we discuss the degree of the delocalization of $f$-electrons
by looking at their contribution to the Fermi surface of the conduction
electrons in our realistic Kondo lattice description for each material.
Even if we have only localized $f$-electrons in our Hamiltonian, they
contribute to the formation of the large Fermi surface via the hybridization
and in this sense they can be delocalized~\cite{martin_1982}.
We follow the procedure used in Ref.~\cite{otsuki_2009_large_fs}.
%
From our simulations for a give material at a fixed temperature,
we get the conduction-electron self energy  $\Sigma(i\omega_{n})$
and then track the temperature dependence of the real part of it
at $i\omega_{n}=0$ to look at the shift of the Fermi level.
Here $\omega_{n}=(2n+1)\pi T$ is the fermion Matsubara frequency. The result is plotted
in Fig.~\ref{plot_minus_ReSigma}. It is seen that CePt$_2$In$_7$ has the most localized
$f$-electrons down to the temperature range of 20~[K] among these materials
while the 115's have the strongest delocalization.
CeIn$_3$ falls in between. We note that the trend in the $f$-electron delocalization
does not exactly follow that in the hybridization strength right on the Fermi level seen
in Fig.~\ref{ImDelta} and it is the outcome of the frequency dependence of the hybridization that we have taken into
account in our realistic Kondo lattice simulations.

To conclude, we have predicted the QCP of CePt$_2$In$_7$
and discussed the possible strategy to have the higher T$_{\rm c}$
in the related material family within our realistic Kondo lattice description
which has been shown to predict the properties of Ce-115's and their parent material
CeIn$_3$ semi-quantitatively.
Our method's predictability of magnetic QCP would further provide the material designing
principle toward more high-$T_{\rm c}$ materials in the upcoming material exploration.

\begin{acknowledgments}
MM thanks Nick Curro, Owen Dix, Ruanchen Dong,
Ravindra Nanguneri, Hiroaki Shishido for discussions.
The present numerical calculations has been done on 
``Chinook'' in Pacific Northwest National Laboratory,
on ``Ranger'' in Texas Advanced Computing Center at the
University of Texas at Austin
under the TeraGrid grant DMR090100, and on a local PC
cluster machine in UC Davis. This work was
supported by DOE SciDAC Grant No. SE-FC02-06ER25793 and by DOE
NEUP Contract No. 00088708.
\end{acknowledgments}



\end{document}